\documentclass{ws-ijmpd}

\begin{document}

\markboth{Hovatta et al.}
{Radio Polarization and gamma-ray emission in AGN jets}

%
\catchline{}{}{}{}{}
%

\title{THE RELATION BETWEEN RADIO POLARIZATION AND GAMMA-RAY EMISSION IN AGN JETS}

\author{TALVIKKI HOVATTA AND MATTHEW L. LISTER}

\address{Department of Physics, Purdue University, 525 Northwestern Ave.\\West Lafayette, IN 47907, USA\\
thovatta@purdue.edu, mlister@purdue.edu}



\author{YURI Y. KOVALEV}

\address{Astro Space Center of Lebedev Physical Institute, Profsoyuznaya 84/32\\117997 Moscow, Russia\\
Max-Planck-Institut f\"ur Radioastronomie, Auf dem H\"ugel 69\\D-53121 Bonn, Germany\\
yyk@asc.rssi.ru}

\author{ALEXANDER B. PUSHKAREV}

\address{Max-Planck-Institut f\"ur Radioastronomie, Auf dem H\"ugel 69\\D-53121 Bonn, Germany\\
Pulkovo Astronomical Observatory, Pulkovskoe Chaussee 65/1\\196140 St. Petersburg, Russia\\
Crimean Astrophysical Observatory\\98688 Nauchny, Crimea, Ukraine\\
apushkar@mpifr.de}

\author{TUOMAS SAVOLAINEN}

\address{Max-Planck-Institut f\"ur Radioastronomie, Auf dem H\"ugel 69\\D-53121 Bonn, Germany\\
tsavolainen@mpifr-bonn.mpg.de}

\maketitle

\begin{history}
\received{Day Month Year}
\revised{Day Month Year}
\comby{Managing Editor}
\end{history}

\begin{abstract}
We have compared the parsec-scale jet linear polarization properties of
the Fermi LAT-detected and non-detected sources in the complete
flux-density-limited (MOJAVE-1) sample of highly beamed AGN. Of the 123
MOJAVE sources, 30 were detected by the LAT during its first three months of
operation. We find that during the era since the launch of Fermi, the unresolved core components of the
LAT-detected jets have significantly higher median fractional polarization at 15
GHz. This complements our previous findings that these LAT sources have
higher apparent jet speeds, brightness temperatures and Doppler factors, and are
preferentially found in higher activity states. 
\end{abstract}

\keywords{Galaxies - active; Galaxies - jets; Quasars - general}

\section{Introduction}	
Active galactic nuclei (AGN) are bright emitters both at radio and $\gamma$-ray wavelengths. In the 1990s the 
Energetic Gamma-Ray Experiment Telescope (EGRET) on board the  {\it Compton Gamma-Ray Observatory} 
detected over 65 AGN (mostly blazars) at $\gamma$-ray energies.\cite{hartman99,mattox01} In June 2008, 
the {\it Fermi Gamma-ray Space Telescope} was launched. Its primary instrument, the Large Area Telescope (LAT), observes 
the whole sky mainly in survey mode at energies 100 MeV-300 GeV.\cite{atwood09} During its first 
three months of operation, the LAT detected 205 bright $\gamma$-ray sources,\cite{abdo09a} of which 
106 were high confidence associations with AGN.\cite{abdo09b} Most of them host bright parsec-scale radio jets.\cite{kovalev09b}
The $\gamma$-ray emission in AGN likely originates in 
the relativistic jet, as suggested by the many correlations found between EGRET and radio/mm observations
of AGN. \cite{valtaoja95}$^-$\cite{kovalev05} Further evidence 
has recently been found using the Very Long Baseline Array (VLBA) observations by Lister et al.\cite{lister09b} who showed 
that the LAT-detected sources have significantly higher apparent jet speeds. It has also been shown that 
the brightness temperature of the VLBI core is higher for LAT-detected sources and the $\gamma$-ray photon 
flux correlates with the compact radio flux density.\cite{kovalev09}
The LAT-detected sources also have larger apparent opening angles,\cite{pushkarev09} and are more Doppler-boosted.\cite{savolainen09}

Possible links between the radio polarization of jets and $\gamma$-ray flaring has been previously studied, 
e.g., by Jorstad et al.\cite{jorstad01a} who established a connection between superluminal component ejections 
and $\gamma$-ray flaring observed by EGRET. Using single-dish radio observations from the University of Michigan 
Radio Astronomy Observatory, they found that a local maximum in the polarized flux density was observed simultaneously with the 
maximum of the $\gamma$-ray emission. Another study by Lister et al.\cite{lister05} compared the 
EGRET detections and 15 GHz VLBI linear polarization properties. They did not find differences between the 
core properties of EGRET-detected and non-detected objects, although they found indications that the EGRET-detected 
objects have higher integrated linear jet polarization. In this paper we will use the better sampled, more complete LAT data 
in the same manner and study how the linear polarization properties of the 15 GHz core differ in the LAT-detected 
and non-detected objects.

\section{The Sample and Data}
In our analyses we use the flux-density-limited MOJAVE-1 (Monitoring Of Jets in Active galactic nuclei with VLBA Experiments) 
sample of highly beamed AGN.\cite{lister09} The sample 
consists of all AGN at declinations $\delta > -20^\circ$ which have had a 15 GHz flux density of at least 1.5 Jy (2 Jy at $\delta < 0^\circ$) 
at any epoch between 1994.0 and 2004.0. We have only used sources at galactic latitude $|b| > 10^\circ$ following
the definitions of the LAT 3-month bright AGN list,\cite{abdo09b} resulting in a sample of 123 sources.
We have divided our sample into LAT-detected (30 objects) and non-LAT-detected (93 objects) sources based on the 
3-month bright AGN list.\cite{abdo09b}

We have used the VLBA 15 GHz total intensity and linear polarization observations since 2002 and determined the polarized core flux 
density $P = (Q^2 + U^2)^{1/2}$, where $Q$ and $U$ are the Stokes parameters, fractional polarization $m_c = P/I$, 
where $I$ is the total flux density of the core, and 
electric vector position angle $\chi = (1/2) \arctan(U/Q)$ at all the epochs where polarization cross-hands were recorded. The core values were defined using the same method as Ref.~\refcite{lister05} by taking the mean over nine contiguous pixels centered at the fitted core position. If the 
polarized flux density was less than five times the rms-value in the $U$ and $Q$ parameters, an upper limit was defined to be five times 
the rms-value. In these cases EVPA was not calculated.

\section{Polarization Levels}
\begin{figure}[pb]
\centerline{\psfig{file=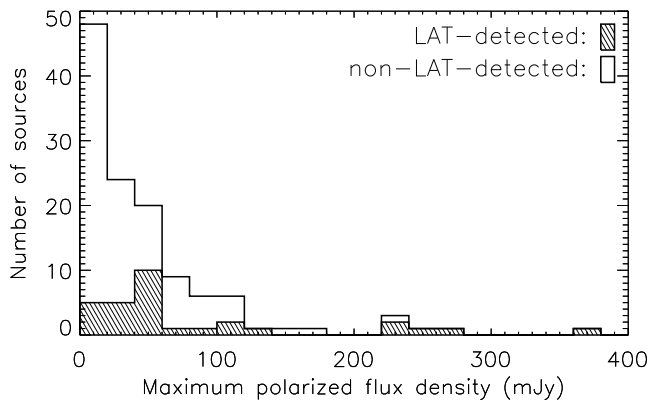,width=9cm}}
\vspace*{8pt}
\caption{Maximum polarized flux density of LAT-detected (shaded) and non-LAT-detected (unshaded) sources since August 2008.\label{f1}}
\end{figure}

In order to study if the polarization properties are connected with $\gamma$-ray activity, we have used 
polarization observations from August 2008 until August 2009 to roughly match the LAT observation period. The LAT associations 
are based on $\gamma$-ray data integrated between August and October 2008.
Figure \ref{f1} shows the distributions of the maximum polarized flux density of LAT-detected and non-LAT-detected sources since August 2008. 
Nine of the the non-LAT-detected sources in the smallest bin are upper limits. The medians of the distributions are 45.0 mJy and 21.6 mJy for the LAT-detected and non-detected sources, respectively. The Gehan's Generalized Wilcoxon test 
in the ASURV package,\cite{lavalley92} suitable for censored data, gives a 
probability p=0.0003 for the distributions of the LAT-detected and non-detected sources to come from the same population. 
The result is not surprising considering that 
the polarized flux density often follows the total flux density which is also seen to be higher in the LAT-detected sources when measured quasi-simultaneously with the LAT.\cite{kovalev09}

\begin{figure}[pb]
\centerline{\psfig{file=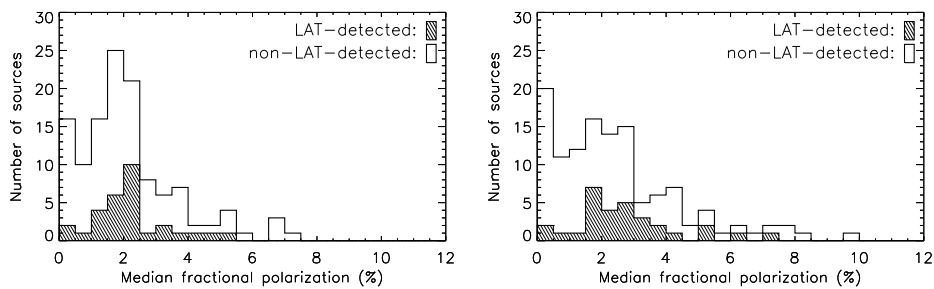,width=\textwidth}}
\vspace*{8pt}
\caption{Median fractional polarization of LAT-detected (shaded) and non-LAT-detected (unshaded) sources from 2002 until August 2008 (left panel) and since August 2008 (right panel).\label{f2}}
\end{figure}

We also wanted to see if this is due to variability in the sources and in Fig.~\ref{f2} we show the distributions of median fractional 
polarization of LAT-detected and non-LAT-detected sources from 2002 until August 2008 and from August 2008 until August 2009 (LAT-era). The median values of the distributions are 2.14\% and 1.80\% for the pre-LAT values of LAT-detected and non-detected sources, 
respectively and 2.50\% and 1.86\% for the values since August 2008. We used the Kolmogorov-Smirnov (K-S) test to see if the LAT-detected 
and non-detected sources come from different populations and the non-parametric Mann-Whitney U-test (M-W U-test) to see if the LAT-detected 
sources have typically higher values than non-detected ones. No significant differences are seen when the pre-LAT values are used (K-S test p=0.35, M-W U-test p=0.16), yet when the distributions from the LAT-era are studied, the groups differ significantly (K-S test p=0.013, M-W U-test p=0.043). 

We calculated the variability indices of fractional polarization (Eq.~\ref{eq1}) and $\chi$ (Eq.~\ref{eq2}) in the method of Refs.~\refcite{aller03} and \refcite{jorstad07}.
\begin{equation}
V_{m_c} = \frac{(m_\mathrm{max} - \sigma_{m_\mathrm{max}}) - (m_\mathrm{min} + \sigma_{m_\mathrm{min}})}{(m_\mathrm{max} - \sigma_{m_\mathrm{max}}) + (m_\mathrm{min} + \sigma_{m_\mathrm{min}})},
\label{eq1}
\end{equation}
where $m_\mathrm{max}$ and $m_\mathrm{min}$ are the maximum and minimum fractional polarization, respectively, measured over all epochs since 2002, and 
$\sigma_{m_\mathrm{max}}$ and $\sigma_{m_\mathrm{min}}$ are the corresponding uncertainties. 
\begin{equation}
V_{\chi} = \frac{|\Delta\chi| - \sqrt{(\sigma_{\chi_{1}}^2 + \sigma_{\chi_{2}}^2)}}{90},
\label{eq2}
\end{equation}
where $|\Delta\chi|$ is the observed range of polarization angle and $\sigma_{\chi_{1}}$ and $\sigma_{\chi_{2}}$ are the uncertainties in the 
two values of EVPAs that define the range.\cite{jorstad07}

\begin{figure}[pb]
\centerline{\psfig{file=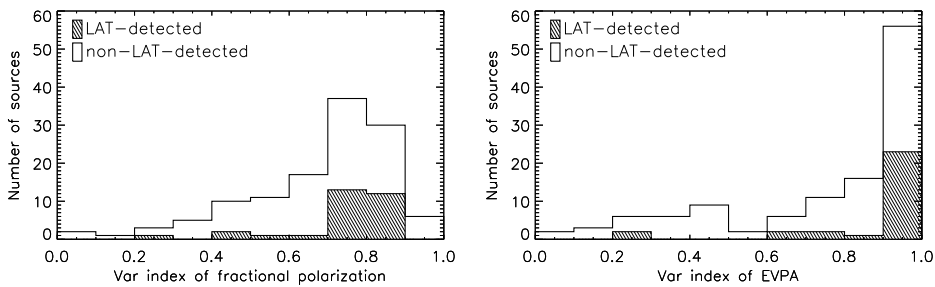,width=\textwidth}}
\vspace*{8pt}
\caption{Variability index of fractional polarization of LAT-detected (shaded) and non-LAT-detected (unshaded) sources from 2002 until August 2008 (left panel) and variability index of EVPA (right panel).\label{f3}}
\end{figure}

Figure \ref{f3} shows the histograms of the variability indices. It can be seen that none of the lowest variability sources are 
detected by the LAT. A K-S test indicates significant differences between the distributions, for both variability indices ($p<0.01$ for $V_{m_c}$ and $p<0.001$ for $V_{\chi}$).
\section{Discussion}
By comparing the median fractional polarization since August 2008 with median values between 2002 and 2008, 
we find the sources to be in a higher polarization state when detected by the LAT.
This complements our previous findings that the LAT-detected sources have higher apparent jet speeds,\cite{lister09b}
higher core brightness temperatures,\cite{kovalev09} larger apparent opening angles\cite{pushkarev09} and larger 
Doppler boosting factors,\cite{savolainen09} In Kovalev et al.\cite{kovalev09} it was also shown that the LAT-detected 
sources are preferentially found in higher radio activity states during the LAT-era. 

The differences seen between detected and non-detected objects can be due to Doppler-beaming.\cite{lister05,lister09b,kovalev09} As is shown in Savolainen et al.\cite{savolainen09}, 
the LAT-detected sources are more boosted, which enhances their observed luminosity more than their non-detected less-beamed counterparts. 
A higher polarization of LAT-detected sources may be explained if the magnetic field in the core is more ordered during the radio / $\gamma$-ray 
flares. Indications that the intrinsic de-aberrated viewing angle in which the source is seen plays a significant role in whether a 
jet is detected at $\gamma$-rays or not, are also seen.\cite{savolainen09} If the radio variations are caused by transverse shocks,\cite{hughes85} the observed fractional polarization can be modeled using the jet rest-frame viewing angle.\cite{hughes85,wardle94}
By assuming that the 15 GHz core is a $\tau = 1$ 
surface and using the jet rest-frame viewing angles from Ref.~\refcite{savolainen09}, we estimated the observed fractional polarization 
for different shock strengths and fractions of uniform magnetic field using simulated values of maximum fractional polarization from 
Ref.~\refcite{homan09} (their Figure A1) and equations of Ref.~\refcite{wardle94}. A uniform shock strength
value for all sources does not reproduce our observations very well. If we let the shock strength vary and choose cases which reproduce the 
observed fractional polarization within 1\%, we see indications that the shocks in BL Lacertae objects 
are stronger than in flat spectrum radio quasars. Further studies are required to confirm our results.  

By using the EGRET observations, Lister et al.\cite{lister05} found no differences in the 15 GHz core polarization properties of $\gamma$-ray-detected and non-detected objects. They only used single-epoch VLBA data while we have used quasi-simultaneous LAT and VLBA
observations. However, the number of LAT-detected sources in our sample is small and therefore we intend to verify our results 
using the LAT 1-year catalog. In this future study we will also examine the polarization of jet components located downstream from the core.

\section*{Acknowledgments}
The authors wish to acknowledge the contributions of the rest of the MOJAVE team. The MOJAVE project is supported under National 
Science Foundation grant AST-0807860 and NASA Fermi grant NNX08AV67G. T.~H. wishes to acknowledge Magnus 
Ehrnrooth Stiftelse for travel funds. Y.~Y.~K. is supported in part by the Russian Foundation for Basic Research
(project 08-02-00545) and a return fellowship of the Alexander von
Humboldt Foundation. T.~S.\ is a research fellow of the Alexander von~Humboldt Foundation. T.~S.\ also acknowledges a support by the Academy of Finland grant 120516. The VLBA is a facility of the National Science Foundation operated by the National
Radio Astronomy Observatory under cooperative agreement with Associated Universities, Inc.

\end{document}